\documentclass[aps,prl,twocolumn,amsmath,amssymb,floatfix,superscriptaddress]{revtex4-1}
\usepackage{tabularx}
\usepackage{bm}
\usepackage{euscript}
\usepackage{epsfig,psfrag,subfigure}
\usepackage{graphicx}
\usepackage{color}
\usepackage{amsfonts}
\usepackage{exscale}
\usepackage{amsbsy}
\usepackage{stmaryrd}
\usepackage{wasysym}

\def\avg#1{\left\langle#1\right\rangle}
\def\bra#1{\left\langle#1\right|}
\def\ket#1{\left|#1\right\rangle}
\def\braket#1#2{\left\langle #1\right|\left.#2\right\rangle}

\def\sgn{{\rm sgn}}

\def\be{\begin{equation}}       \def\ee{\end{equation}}
\def\bea{\begin{eqnarray}}      \def\eea{\end{eqnarray}}
\def\ba{\begin{array} }
\def\ea{\end{array} }
\def\bnum{\begin{enumerate} }
\def\enum{\end{enumerate}}

\def\nn{\nonumber}

\def\=>{\Rightarrow}
\def\>{\rightarrow}

\def\eye2{Fathbb{I}}

\def\Eq#1{Eq.~(\ref{#1})}
\def\Fig#1{Fig.~\ref{#1}}

\begin{document}
\title{Exact Spin Liquid Ground States of the Quantum Dimer Model on the Square and Honeycomb Lattices}
\author{Hong Yao}
\affiliation{Department of Physics, Stanford University, Stanford, CA 94305, USA}
\affiliation{Institute for Advanced Study, Tsinghua University, Beijing, 100084, China}
\author{Steven A. Kivelson}
\affiliation{Department of Physics, Stanford University, Stanford, CA 94305, USA}
\date{\today}
\begin{abstract}
We study a generalized quantum hard-core dimer model on the square and honeycomb lattices, allowing for first and second neighbor dimers. At generalized RK points, the exact ground states can be constructed, and ground-state correlation functions can be equated to those of {\it interacting} 1+1 dimensional Grassmann fields. When the concentration of second neighbor dimers is small, the ground state correlations are shown to be short-ranged corresponding to a (gaped) spin liquid phase. On a 2-torus, the ground states exhibit fourfold topological degeneracy. On a finite cylinder we have found a dramatic even-odd effect depending on the circumference, and propose that this can be used as a numerical diagnostic of gapped spin-liquid phases, more generally.
\end{abstract}

\maketitle
There has been a surge in numerical experiments \cite{whitehuse,assaad,yao} on simple
models reporting evidence of the existence of fully gapped (short-ranged) spin-liquid states, including reports for the spin-$1/2$ Heisenberg model with first and second neighbor anti-ferromagnetic
couplings ($J_1$ and $J_2$) on the square \cite{yao} and honeycomb \cite{clark} lattice as well as the Hubbard model on the
honeycomb lattice\cite{assaad}. For the square lattice model, an early numerical study\cite{figurdido} on a relatively small system was interpreted as evidence for a spin-liquid, but this result was seriously questioned on the basis of the results of series expansion\cite{singh}, large-$N$\cite{sachdevread}, and large-$S$\cite{larkin} studies of the same model. In the case of the honeycomb lattice,
recent studies reported\cite{assaad} evidence of such a state for an intermediate range of $U/t$ in the Hubbard model.  For both the square and honeycomb lattice, there is no
solvable model with the same symmetries as the Heisenberg model for which a gaped spin-liquid state has been found.  Having %such
a caricature of the putative state is useful in thinking about the phase, and possibly also in designing %future
numerical experiments to confirm or falsify its existence.

Here, we report a
generalization of the quantum hard-core dimer model\cite{rk,moessnersondhi} on the square and honeycomb lattices, for which the exact spin-liquid groundstates can be obtained for certain values of coupling constants dubbed as ``generalized Rokhsar-Kivelson (RK) points''. In the limit in which all dimers occupy nearest-neighbor links, this spin-liquid
has critical ground-state correlations, and correspondingly gapless collective excitations\cite{rk,ardonne,fradkin}.  However, when the dynamics are generalized to generate even a small concentration of second neighbor dimers, the ground-state correlations become short-ranged and a gap opens in the spectrum.  Indeed, although the ground-state correlations in this limit can no longer be computed directly using Pfaffian methods\cite{pfaffian}, close to the critical point, where the concentration of second neighbor dimers is small, the correlations are asymptotically equivalent to those of a 1+1 dimensional massive Thirring model, and so are still exactly known\cite{thirring}.

In the course of this study, we have also recognized a new, possibly useful diagnostic tool for numerical searches for spin-liquids:  We find the exact spin-liquid ground state
on a torus with a finite circumference, $L_y$,
spontaneously breaks translational symmetry (forms a ``columnar'' density wave state) for odd $L_y$, but is translationally invariant for even $L_y$.
(A similar
phenomenon occurs when a fractional quantum Hall fluid is considered in the narrow torus limit\cite{karlhede,dhlee}.)
For odd $L_y$, the amplitude of the columnar order parameter decays exponentially in proportion to $\frac{1}{L_y^{1/2}}\exp[-L_y/2\xi]$,
where $\xi$ is
the dimer-dimer correlation length\cite{footnote}.  We note that a similar even-odd effect with exponential decrease in the amplitude of the columnar order parameter for odd $L_y$ have been observed\cite{yao} in DMRG studies of the square-lattice spin-1/2 $J_1$-$J_2$ model on cylinders with
 $L_y=3-10$.

{\bf Model:}  The quantum dimer model is defined on a Hilbert space with distinct, orthonormal states corresponding to each allowed hard-core dimer covering of the lattice.  The Hamiltonian is defined by matrix elements between dimer states, with ``potential'' terms, $V,\ V^\prime,\ \lambda V^{\prime\prime}$, and $V^{\prime\prime}/\lambda$, which are diagonal in the dimer basis and associate an interaction energy with various local arrangements of dimers, and  ``kinetic'' terms, $t,\ t^\prime$, and $t^{\prime\prime}$, which involve a local rearrangement of a small number of dimers.  On the square lattice, we represent the Hamiltonian graphically as:
\bea\label{eq:square}
&&
H=\sum_{\square}\Big[\!-\!t\left(\ket{% 
%       2         4        6
%       -         -
%       -         -
%       -         -
%       1         3        5
%
\setlength{\unitlength}{3947sp}
\begin{picture}(160,120)(-20,10)
 \put (0,0){\circle*{25}}
 \put (120,0){\circle*{25}}
 \put (0,120){\circle*{25}}
 \put (120,120){\circle*{25}}
 
 %\put (0,0){\line(0,1){120}} 
 %\put (120,0){\line(0,1){120}}   
 \put (0,10){\circle*{1}}
 \put (0,20){\circle*{1}}
 \put (0,30){\circle*{1}}
 \put (0,40){\circle*{1}}    
 \put (0,50){\circle*{1}}
 \put (0,60){\circle*{1}}  
 \put (0,70){\circle*{1}}
 \put (0,80){\circle*{1}}    
 \put (0,90){\circle*{1}}
 \put (0,100){\circle*{1}}    
 \put (0,110){\circle*{1}}
 
 \put (120,10){\circle*{1}}
 \put (120,20){\circle*{1}}
 \put (120,30){\circle*{1}}
 \put (120,40){\circle*{1}}    
 \put (120,50){\circle*{1}}
 \put (120,60){\circle*{1}}  
 \put (120,70){\circle*{1}}
 \put (120,80){\circle*{1}}    
 \put (120,90){\circle*{1}}
 \put (120,100){\circle*{1}}    
 \put (120,110){\circle*{1}} 
\end{picture}} \bra{% 
%       2--------4        6
%
%          
%
%       1--------3        5
%
\setlength{\unitlength}{3947sp}
\begin{picture}(160,120)(-20,10)
 \put (0,0){\circle*{25}}
 \put (120,0){\circle*{25}}
 \put (0,120){\circle*{25}}
 \put (120,120){\circle*{25}}
 %\put (0,120){\line(1,0){120}} 
 %\put (0,0){\line(1,0){120}}  

 \put (10,0){\circle*{1}}
 \put (20,0){\circle*{1}}
 \put (30,0){\circle*{1}}
 \put (40,0){\circle*{1}}    
 \put (50,0){\circle*{1}}
 \put (60,0){\circle*{1}}  
 \put (70,0){\circle*{1}}
 \put (80,0){\circle*{1}}    
 \put (90,0){\circle*{1}}
 \put (100,0){\circle*{1}}    
 \put (110,0){\circle*{1}}
 
 \put (10,120){\circle*{1}}
 \put (20,120){\circle*{1}}
 \put (30,120){\circle*{1}}
 \put (40,120){\circle*{1}}    
 \put (50,120){\circle*{1}}
 \put (60,120){\circle*{1}}  
 \put (70,120){\circle*{1}}
 \put (80,120){\circle*{1}}    
 \put (90,120){\circle*{1}}
 \put (100,120){\circle*{1}}    
 \put (110,120){\circle*{1}} 
    
\end{picture}} \!+\!h.c.\right) \!+\!V\left(\ket{% 
%       2         4        6
%       -         -
%       -         -
%       -         -
%       1         3        5
%
\setlength{\unitlength}{3947sp}
\begin{picture}(160,120)(-20,10)
 \put (0,0){\circle*{25}}
 \put (120,0){\circle*{25}}
 \put (0,120){\circle*{25}}
 \put (120,120){\circle*{25}}
 
 %\put (0,0){\line(0,1){120}} 
 %\put (120,0){\line(0,1){120}}   
 \put (0,10){\circle*{1}}
 \put (0,20){\circle*{1}}
 \put (0,30){\circle*{1}}
 \put (0,40){\circle*{1}}    
 \put (0,50){\circle*{1}}
 \put (0,60){\circle*{1}}  
 \put (0,70){\circle*{1}}
 \put (0,80){\circle*{1}}    
 \put (0,90){\circle*{1}}
 \put (0,100){\circle*{1}}    
 \put (0,110){\circle*{1}}
 
 \put (120,10){\circle*{1}}
 \put (120,20){\circle*{1}}
 \put (120,30){\circle*{1}}
 \put (120,40){\circle*{1}}    
 \put (120,50){\circle*{1}}
 \put (120,60){\circle*{1}}  
 \put (120,70){\circle*{1}}
 \put (120,80){\circle*{1}}    
 \put (120,90){\circle*{1}}
 \put (120,100){\circle*{1}}    
 \put (120,110){\circle*{1}} 
\end{picture}}\bra{% 
%       2         4        6
%       -         -
%       -         -
%       -         -
%       1         3        5
%
\setlength{\unitlength}{3947sp}
\begin{picture}(160,120)(-20,10)
 \put (0,0){\circle*{25}}
 \put (120,0){\circle*{25}}
 \put (0,120){\circle*{25}}
 \put (120,120){\circle*{25}}
 
 %\put (0,0){\line(0,1){120}} 
 %\put (120,0){\line(0,1){120}}   
 \put (0,10){\circle*{1}}
 \put (0,20){\circle*{1}}
 \put (0,30){\circle*{1}}
 \put (0,40){\circle*{1}}    
 \put (0,50){\circle*{1}}
 \put (0,60){\circle*{1}}  
 \put (0,70){\circle*{1}}
 \put (0,80){\circle*{1}}    
 \put (0,90){\circle*{1}}
 \put (0,100){\circle*{1}}    
 \put (0,110){\circle*{1}}
 
 \put (120,10){\circle*{1}}
 \put (120,20){\circle*{1}}
 \put (120,30){\circle*{1}}
 \put (120,40){\circle*{1}}    
 \put (120,50){\circle*{1}}
 \put (120,60){\circle*{1}}  
 \put (120,70){\circle*{1}}
 \put (120,80){\circle*{1}}    
 \put (120,90){\circle*{1}}
 \put (120,100){\circle*{1}}    
 \put (120,110){\circle*{1}} 
\end{picture}} +\ket{% 
%       2--------4        6
%
%          
%
%       1--------3        5
%
\setlength{\unitlength}{3947sp}
\begin{picture}(160,120)(-20,10)
 \put (0,0){\circle*{25}}
 \put (120,0){\circle*{25}}
 \put (0,120){\circle*{25}}
 \put (120,120){\circle*{25}}
 %\put (0,120){\line(1,0){120}} 
 %\put (0,0){\line(1,0){120}}  

 \put (10,0){\circle*{1}}
 \put (20,0){\circle*{1}}
 \put (30,0){\circle*{1}}
 \put (40,0){\circle*{1}}    
 \put (50,0){\circle*{1}}
 \put (60,0){\circle*{1}}  
 \put (70,0){\circle*{1}}
 \put (80,0){\circle*{1}}    
 \put (90,0){\circle*{1}}
 \put (100,0){\circle*{1}}    
 \put (110,0){\circle*{1}}
 
 \put (10,120){\circle*{1}}
 \put (20,120){\circle*{1}}
 \put (30,120){\circle*{1}}
 \put (40,120){\circle*{1}}    
 \put (50,120){\circle*{1}}
 \put (60,120){\circle*{1}}  
 \put (70,120){\circle*{1}}
 \put (80,120){\circle*{1}}    
 \put (90,120){\circle*{1}}
 \put (100,120){\circle*{1}}    
 \put (110,120){\circle*{1}} 
    
\end{picture}}\bra{% 
%       2--------4        6
%
%          
%
%       1--------3        5
%
\setlength{\unitlength}{3947sp}
\begin{picture}(160,120)(-20,10)
 \put (0,0){\circle*{25}}
 \put (120,0){\circle*{25}}
 \put (0,120){\circle*{25}}
 \put (120,120){\circle*{25}}
 %\put (0,120){\line(1,0){120}} 
 %\put (0,0){\line(1,0){120}}  

 \put (10,0){\circle*{1}}
 \put (20,0){\circle*{1}}
 \put (30,0){\circle*{1}}
 \put (40,0){\circle*{1}}    
 \put (50,0){\circle*{1}}
 \put (60,0){\circle*{1}}  
 \put (70,0){\circle*{1}}
 \put (80,0){\circle*{1}}    
 \put (90,0){\circle*{1}}
 \put (100,0){\circle*{1}}    
 \put (110,0){\circle*{1}}
 
 \put (10,120){\circle*{1}}
 \put (20,120){\circle*{1}}
 \put (30,120){\circle*{1}}
 \put (40,120){\circle*{1}}    
 \put (50,120){\circle*{1}}
 \put (60,120){\circle*{1}}  
 \put (70,120){\circle*{1}}
 \put (80,120){\circle*{1}}    
 \put (90,120){\circle*{1}}
 \put (100,120){\circle*{1}}    
 \put (110,120){\circle*{1}} 
    
\end{picture}}\right)\Big]\nn
\\
&&\!+\!\sum_{\{% 
%       2        4        6
%             -        -
%           -        -
%         -        -
%       1        3        5
%
\setlength{\unitlength}{2750sp}
\begin{picture}(89,160)(-5,20)
 \put (0,0){\line(1,0){80}}
 \put (0,80){\line(1,0){80}}
 \put (0,160){\line(1,0){80}}
  
 \put (0,0){\line(0,1){160}}
 \put (80,0){\line(0,1){160}}
\end{picture}
\}} \Big[\!-\!t'\left(\ket{% 
%       2        4        6
%                  -   
%                    -
%                      -
%       1 - - -  3        5
%
\setlength{\unitlength}{3947sp}
\begin{picture}(130,160)(-15,60)
 \put (0,0){\circle*{25}}
 \put (100,0){\circle*{25}}
 \put (0,100){\circle*{25}}
 \put (100,100){\circle*{25}}
 \put (0,200){\circle*{25}}
 \put (100,200){\circle*{25}}
 
 %\put (0,0){\line(0,1){100}} 

 \put (0,10){\circle*{1}}
 \put (0,20){\circle*{1}}
 \put (0,30){\circle*{1}}
 \put (0,40){\circle*{1}}    
 \put (0,50){\circle*{1}}
 \put (0,60){\circle*{1}}  
 \put (0,70){\circle*{1}}
 \put (0,80){\circle*{1}}    
 \put (0,90){\circle*{1}}

 %45
 \put (10,190){\circle*{1}}
 \put (20,180){\circle*{1}}
 \put (30,170){\circle*{1}}
 \put (40,160){\circle*{1}}    
 \put (50,150){\circle*{1}}
 \put (60,140){\circle*{1}}  
 \put (70,130){\circle*{1}}
 \put (80,120){\circle*{1}}    
 \put (90,110){\circle*{1}}

\end{picture}}\bra{% 
%       2        4        6
%                  -   
%                    -
%                      -
%       1 - - -  3        5
%
\setlength{\unitlength}{3947sp}
\begin{picture}(130,160)(-15,60)
 \put (0,0){\circle*{25}}
 \put (100,0){\circle*{25}}
 \put (0,100){\circle*{25}}
 \put (100,100){\circle*{25}}
 \put (0,200){\circle*{25}}
 \put (100,200){\circle*{25}}
 
 %\put (0,100){\line(0,1){100}}  

\put (0,110){\circle*{1}}
 \put (0,120){\circle*{1}}
 \put (0,130){\circle*{1}}
 \put (0,140){\circle*{1}}    
 \put (0,150){\circle*{1}}
 \put (0,160){\circle*{1}}  
 \put (0,170){\circle*{1}}
 \put (0,180){\circle*{1}}    
 \put (0,190){\circle*{1}}

 %14
 \put (10,10){\circle*{1}}
 \put (20,20){\circle*{1}}
 \put (30,30){\circle*{1}}
 \put (40,40){\circle*{1}}    
 \put (50,50){\circle*{1}}
 \put (60,60){\circle*{1}}  
 \put (70,70){\circle*{1}}
 \put (80,80){\circle*{1}}    
 \put (90,90){\circle*{1}}
    
\end{picture}} +h.c.\right)\!+\!V'\left(\ket{% 
%       2        4        6
%                  -   
%                    -
%                      -
%       1 - - -  3        5
%
\setlength{\unitlength}{3947sp}
\begin{picture}(130,160)(-15,60)
 \put (0,0){\circle*{25}}
 \put (100,0){\circle*{25}}
 \put (0,100){\circle*{25}}
 \put (100,100){\circle*{25}}
 \put (0,200){\circle*{25}}
 \put (100,200){\circle*{25}}
 
 %\put (0,0){\line(0,1){100}} 

 \put (0,10){\circle*{1}}
 \put (0,20){\circle*{1}}
 \put (0,30){\circle*{1}}
 \put (0,40){\circle*{1}}    
 \put (0,50){\circle*{1}}
 \put (0,60){\circle*{1}}  
 \put (0,70){\circle*{1}}
 \put (0,80){\circle*{1}}    
 \put (0,90){\circle*{1}}

 %45
 \put (10,190){\circle*{1}}
 \put (20,180){\circle*{1}}
 \put (30,170){\circle*{1}}
 \put (40,160){\circle*{1}}    
 \put (50,150){\circle*{1}}
 \put (60,140){\circle*{1}}  
 \put (70,130){\circle*{1}}
 \put (80,120){\circle*{1}}    
 \put (90,110){\circle*{1}}

\end{picture}} \bra{% 
%       2        4        6
%                  -   
%                    -
%                      -
%       1 - - -  3        5
%
\setlength{\unitlength}{3947sp}
\begin{picture}(130,160)(-15,60)
 \put (0,0){\circle*{25}}
 \put (100,0){\circle*{25}}
 \put (0,100){\circle*{25}}
 \put (100,100){\circle*{25}}
 \put (0,200){\circle*{25}}
 \put (100,200){\circle*{25}}
 
 %\put (0,0){\line(0,1){100}} 

 \put (0,10){\circle*{1}}
 \put (0,20){\circle*{1}}
 \put (0,30){\circle*{1}}
 \put (0,40){\circle*{1}}    
 \put (0,50){\circle*{1}}
 \put (0,60){\circle*{1}}  
 \put (0,70){\circle*{1}}
 \put (0,80){\circle*{1}}    
 \put (0,90){\circle*{1}}

 %45
 \put (10,190){\circle*{1}}
 \put (20,180){\circle*{1}}
 \put (30,170){\circle*{1}}
 \put (40,160){\circle*{1}}    
 \put (50,150){\circle*{1}}
 \put (60,140){\circle*{1}}  
 \put (70,130){\circle*{1}}
 \put (80,120){\circle*{1}}    
 \put (90,110){\circle*{1}}

\end{picture}} +\ket{% 
%       2        4        6
%                  -   
%                    -
%                      -
%       1 - - -  3        5
%
\setlength{\unitlength}{3947sp}
\begin{picture}(130,160)(-15,60)
 \put (0,0){\circle*{25}}
 \put (100,0){\circle*{25}}
 \put (0,100){\circle*{25}}
 \put (100,100){\circle*{25}}
 \put (0,200){\circle*{25}}
 \put (100,200){\circle*{25}}
 
 %\put (0,100){\line(0,1){100}}  

\put (0,110){\circle*{1}}
 \put (0,120){\circle*{1}}
 \put (0,130){\circle*{1}}
 \put (0,140){\circle*{1}}    
 \put (0,150){\circle*{1}}
 \put (0,160){\circle*{1}}  
 \put (0,170){\circle*{1}}
 \put (0,180){\circle*{1}}    
 \put (0,190){\circle*{1}}

 %14
 \put (10,10){\circle*{1}}
 \put (20,20){\circle*{1}}
 \put (30,30){\circle*{1}}
 \put (40,40){\circle*{1}}    
 \put (50,50){\circle*{1}}
 \put (60,60){\circle*{1}}  
 \put (70,70){\circle*{1}}
 \put (80,80){\circle*{1}}    
 \put (90,90){\circle*{1}}
    
\end{picture}}\bra{% 
%       2        4        6
%                  -   
%                    -
%                      -
%       1 - - -  3        5
%
\setlength{\unitlength}{3947sp}
\begin{picture}(130,160)(-15,60)
 \put (0,0){\circle*{25}}
 \put (100,0){\circle*{25}}
 \put (0,100){\circle*{25}}
 \put (100,100){\circle*{25}}
 \put (0,200){\circle*{25}}
 \put (100,200){\circle*{25}}
 
 %\put (0,100){\line(0,1){100}}  

\put (0,110){\circle*{1}}
 \put (0,120){\circle*{1}}
 \put (0,130){\circle*{1}}
 \put (0,140){\circle*{1}}    
 \put (0,150){\circle*{1}}
 \put (0,160){\circle*{1}}  
 \put (0,170){\circle*{1}}
 \put (0,180){\circle*{1}}    
 \put (0,190){\circle*{1}}

 %14
 \put (10,10){\circle*{1}}
 \put (20,20){\circle*{1}}
 \put (30,30){\circle*{1}}
 \put (40,40){\circle*{1}}    
 \put (50,50){\circle*{1}}
 \put (60,60){\circle*{1}}  
 \put (70,70){\circle*{1}}
 \put (80,80){\circle*{1}}    
 \put (90,90){\circle*{1}}
    
\end{picture}}\right)  \nn\\
&&-t^{\prime\prime}\left(\ket{% 
%       2        4--------6
%
%             
%
%       1--------3        5
%
\setlength{\unitlength}{3947sp}
\begin{picture}(130,160)(-15,60)
 \put (0,0){\circle*{25}}
 \put (100,0){\circle*{25}}
 \put (0,100){\circle*{25}}
 \put (100,100){\circle*{25}}
 \put (0,200){\circle*{25}}
 \put (100,200){\circle*{25}}
 
 %\put (0,0){\line(0,1){100}} 
 %\put (100,100){\line(0,1){100}}    
 
 \put (0,10){\circle*{1}}
 \put (0,20){\circle*{1}}
 \put (0,30){\circle*{1}}
 \put (0,40){\circle*{1}}    
 \put (0,50){\circle*{1}}
 \put (0,60){\circle*{1}}  
 \put (0,70){\circle*{1}}
 \put (0,80){\circle*{1}}    
 \put (0,90){\circle*{1}}
 
 \put (100,110){\circle*{1}}
 \put (100,120){\circle*{1}}
 \put (100,130){\circle*{1}}
 \put (100,140){\circle*{1}}    
 \put (100,150){\circle*{1}}
 \put (100,160){\circle*{1}}  
 \put (100,170){\circle*{1}}
 \put (100,180){\circle*{1}}    
 \put (100,190){\circle*{1}}

\end{picture}} \bra{% 
%       2        4        6
%             -        -
%           -        -
%         -        -
%       1        3        5
%
\setlength{\unitlength}{3947sp}
\begin{picture}(130,160)(-15,60)
 \put (0,0){\circle*{25}}
 \put (100,0){\circle*{25}}
 \put (0,100){\circle*{25}}
 \put (100,100){\circle*{25}}
 \put (0,200){\circle*{25}}
 \put (100,200){\circle*{25}}
 
 \put (10,10){\circle*{1}}
 \put (20,20){\circle*{1}}
 \put (30,30){\circle*{1}}
 \put (40,40){\circle*{1}}    
 \put (50,50){\circle*{1}}
 \put (60,60){\circle*{1}}  
 \put (70,70){\circle*{1}}
 \put (80,80){\circle*{1}}    
 \put (90,90){\circle*{1}}
 \put (95,95){\circle*{1}}

 \put (10,110){\circle*{1}}
 \put (20,120){\circle*{1}}
 \put (30,130){\circle*{1}}
 \put (40,140){\circle*{1}}    
 \put (50,150){\circle*{1}}
 \put (60,160){\circle*{1}}  
 \put (70,170){\circle*{1}}
 \put (80,180){\circle*{1}}    
 \put (90,190){\circle*{1}}
 \put (95,195){\circle*{1}}
 
\end{picture}} \!+\!h.c.\right)\!+\! \lambda V^{\prime\prime}\ket{% 
%       2        4--------6
%
%             
%
%       1--------3        5
%
\setlength{\unitlength}{3947sp}
\begin{picture}(130,160)(-15,60)
 \put (0,0){\circle*{25}}
 \put (100,0){\circle*{25}}
 \put (0,100){\circle*{25}}
 \put (100,100){\circle*{25}}
 \put (0,200){\circle*{25}}
 \put (100,200){\circle*{25}}
 
 %\put (0,0){\line(0,1){100}} 
 %\put (100,100){\line(0,1){100}}    
 
 \put (0,10){\circle*{1}}
 \put (0,20){\circle*{1}}
 \put (0,30){\circle*{1}}
 \put (0,40){\circle*{1}}    
 \put (0,50){\circle*{1}}
 \put (0,60){\circle*{1}}  
 \put (0,70){\circle*{1}}
 \put (0,80){\circle*{1}}    
 \put (0,90){\circle*{1}}
 
 \put (100,110){\circle*{1}}
 \put (100,120){\circle*{1}}
 \put (100,130){\circle*{1}}
 \put (100,140){\circle*{1}}    
 \put (100,150){\circle*{1}}
 \put (100,160){\circle*{1}}  
 \put (100,170){\circle*{1}}
 \put (100,180){\circle*{1}}    
 \put (100,190){\circle*{1}}

\end{picture}}\bra{% 
%       2        4--------6
%
%             
%
%       1--------3        5
%
\setlength{\unitlength}{3947sp}
\begin{picture}(130,160)(-15,60)
 \put (0,0){\circle*{25}}
 \put (100,0){\circle*{25}}
 \put (0,100){\circle*{25}}
 \put (100,100){\circle*{25}}
 \put (0,200){\circle*{25}}
 \put (100,200){\circle*{25}}
 
 %\put (0,0){\line(0,1){100}} 
 %\put (100,100){\line(0,1){100}}    
 
 \put (0,10){\circle*{1}}
 \put (0,20){\circle*{1}}
 \put (0,30){\circle*{1}}
 \put (0,40){\circle*{1}}    
 \put (0,50){\circle*{1}}
 \put (0,60){\circle*{1}}  
 \put (0,70){\circle*{1}}
 \put (0,80){\circle*{1}}    
 \put (0,90){\circle*{1}}
 
 \put (100,110){\circle*{1}}
 \put (100,120){\circle*{1}}
 \put (100,130){\circle*{1}}
 \put (100,140){\circle*{1}}    
 \put (100,150){\circle*{1}}
 \put (100,160){\circle*{1}}  
 \put (100,170){\circle*{1}}
 \put (100,180){\circle*{1}}    
 \put (100,190){\circle*{1}}

\end{picture}} \!+\!\frac{V^{\prime\prime}}{\lambda} \ket{% 
%       2        4        6
%             -        -
%           -        -
%         -        -
%       1        3        5
%
\setlength{\unitlength}{3947sp}
\begin{picture}(130,160)(-15,60)
 \put (0,0){\circle*{25}}
 \put (100,0){\circle*{25}}
 \put (0,100){\circle*{25}}
 \put (100,100){\circle*{25}}
 \put (0,200){\circle*{25}}
 \put (100,200){\circle*{25}}
 
 \put (10,10){\circle*{1}}
 \put (20,20){\circle*{1}}
 \put (30,30){\circle*{1}}
 \put (40,40){\circle*{1}}    
 \put (50,50){\circle*{1}}
 \put (60,60){\circle*{1}}  
 \put (70,70){\circle*{1}}
 \put (80,80){\circle*{1}}    
 \put (90,90){\circle*{1}}
 \put (95,95){\circle*{1}}

 \put (10,110){\circle*{1}}
 \put (20,120){\circle*{1}}
 \put (30,130){\circle*{1}}
 \put (40,140){\circle*{1}}    
 \put (50,150){\circle*{1}}
 \put (60,160){\circle*{1}}  
 \put (70,170){\circle*{1}}
 \put (80,180){\circle*{1}}    
 \put (90,190){\circle*{1}}
 \put (95,195){\circle*{1}}
 
\end{picture}}\bra{% 
%       2        4        6
%             -        -
%           -        -
%         -        -
%       1        3        5
%
\setlength{\unitlength}{3947sp}
\begin{picture}(130,160)(-15,60)
 \put (0,0){\circle*{25}}
 \put (100,0){\circle*{25}}
 \put (0,100){\circle*{25}}
 \put (100,100){\circle*{25}}
 \put (0,200){\circle*{25}}
 \put (100,200){\circle*{25}}
 
 \put (10,10){\circle*{1}}
 \put (20,20){\circle*{1}}
 \put (30,30){\circle*{1}}
 \put (40,40){\circle*{1}}    
 \put (50,50){\circle*{1}}
 \put (60,60){\circle*{1}}  
 \put (70,70){\circle*{1}}
 \put (80,80){\circle*{1}}    
 \put (90,90){\circle*{1}}
 \put (95,95){\circle*{1}}

 \put (10,110){\circle*{1}}
 \put (20,120){\circle*{1}}
 \put (30,130){\circle*{1}}
 \put (40,140){\circle*{1}}    
 \put (50,150){\circle*{1}}
 \put (60,160){\circle*{1}}  
 \put (70,170){\circle*{1}}
 \put (80,180){\circle*{1}}    
 \put (90,190){\circle*{1}}
 \put (95,195){\circle*{1}}
 
\end{picture}}\Big],
\eea
where black bonds are occupied by dimers, the first sum runs over all plaquettes, and the second and third over all pairs of adjacent plaquettes (with both orientations). The first line is \Eq{eq:square} is the original quantum dimer model
on the square lattice with only first neighbor dimers\cite{rk}.  The added terms are the shortest-range terms involving second-neighbor dimers.  The parameter $\lambda$ determines the relative preference for first and second-neighbor dimers.

A similar construction can be used to define the model on the honeycomb lattice:
\bea\label{eq:honeycomb}
&&\!\!
H\!=\!\sum_{\varhexagon}\!\Big[\!-\!
t\left(\ket{%
%         /  5
%       6       4
%               |        
%       1       3
%         \  2      
%
\setlength{\unitlength}{3947sp}
\begin{picture}(160,150)(-10,30)
 \put (0,40){\circle*{25}}
 \put (70,0){\circle*{25}}
 \put (140,40){\circle*{25}}
 \put (140,120){\circle*{25}}
 \put (70,160){\circle*{25}}
 \put (0,120){\circle*{25}}

%34
% \put (140,40){\line(0,1){80}} 
 \put (140,50){\circle*{1}}
 \put (140,60){\circle*{1}}
 \put (140,70){\circle*{1}}  
 \put (140,80){\circle*{1}}
 \put (140,90){\circle*{1}}
 \put (140,100){\circle*{1}}
 \put (140,110){\circle*{1}}
 
%12 
 \put (7,36){\circle*{1}}
 \put (14,32){\circle*{1}}
 \put (21,28){\circle*{1}}
 \put (28,24){\circle*{1}}    
 \put (35,20){\circle*{1}}
 \put (42,16){\circle*{1}}  
 \put (49,12){\circle*{1}}
 \put (56,8){\circle*{1}}
 \put (63,4){\circle*{1}}

 %56
 \put (7,124){\circle*{1}}
 \put (14,128){\circle*{1}}
 \put (21,132){\circle*{1}}
 \put (28,136){\circle*{1}}    
 \put (35,140){\circle*{1}}
 \put (42,144){\circle*{1}}  
 \put (49,148){\circle*{1}}
 \put (56,152){\circle*{1}}
 \put (63,156){\circle*{1}}

\end{picture}} \bra{%
%           5\
%       6       4
%       |
%       1       3
%           2 /     
%
\setlength{\unitlength}{3947sp}
\begin{picture}(160,150)(-10,30)
 \put (0,40){\circle*{25}}
 \put (70,0){\circle*{25}}
 \put (140,40){\circle*{25}}
 \put (140,120){\circle*{25}}
 \put (70,160){\circle*{25}}
 \put (0,120){\circle*{25}}
 
 %\put (0,40){\line(0,1){80}}
  \put (0,60){\circle*{1}} 
  \put (0,70){\circle*{1}}
  \put (0,80){\circle*{1}}
  \put (0,90){\circle*{1}}  
  \put (0,100){\circle*{1}}
  \put (0,110){\circle*{1}}

 \put (133,36){\circle*{1}}
 \put (126,32){\circle*{1}}
 \put (119,28){\circle*{1}}
 \put (112,24){\circle*{1}}    
 \put (105,20){\circle*{1}}
 \put (98,16){\circle*{1}}  
 \put (91,12){\circle*{1}}
 \put (84,8){\circle*{1}}
 \put (77,4){\circle*{1}}

 \put (133,124){\circle*{1}}
 \put (126,128){\circle*{1}}
 \put (119,132){\circle*{1}}
 \put (112,136){\circle*{1}}    
 \put (105,140){\circle*{1}}
 \put (98,144){\circle*{1}}  
 \put (91,148){\circle*{1}}
 \put (84,152){\circle*{1}}
 \put (77,156){\circle*{1}}   
\end{picture}} +h.c.\right) \!+\!V\left(\ket{%
%         /  5
%       6       4
%               |        
%       1       3
%         \  2      
%
\setlength{\unitlength}{3947sp}
\begin{picture}(160,150)(-10,30)
 \put (0,40){\circle*{25}}
 \put (70,0){\circle*{25}}
 \put (140,40){\circle*{25}}
 \put (140,120){\circle*{25}}
 \put (70,160){\circle*{25}}
 \put (0,120){\circle*{25}}

%34
% \put (140,40){\line(0,1){80}} 
 \put (140,50){\circle*{1}}
 \put (140,60){\circle*{1}}
 \put (140,70){\circle*{1}}  
 \put (140,80){\circle*{1}}
 \put (140,90){\circle*{1}}
 \put (140,100){\circle*{1}}
 \put (140,110){\circle*{1}}
 
%12 
 \put (7,36){\circle*{1}}
 \put (14,32){\circle*{1}}
 \put (21,28){\circle*{1}}
 \put (28,24){\circle*{1}}    
 \put (35,20){\circle*{1}}
 \put (42,16){\circle*{1}}  
 \put (49,12){\circle*{1}}
 \put (56,8){\circle*{1}}
 \put (63,4){\circle*{1}}

 %56
 \put (7,124){\circle*{1}}
 \put (14,128){\circle*{1}}
 \put (21,132){\circle*{1}}
 \put (28,136){\circle*{1}}    
 \put (35,140){\circle*{1}}
 \put (42,144){\circle*{1}}  
 \put (49,148){\circle*{1}}
 \put (56,152){\circle*{1}}
 \put (63,156){\circle*{1}}

\end{picture}}\bra{%
%         /  5
%       6       4
%               |        
%       1       3
%         \  2      
%
\setlength{\unitlength}{3947sp}
\begin{picture}(160,150)(-10,30)
 \put (0,40){\circle*{25}}
 \put (70,0){\circle*{25}}
 \put (140,40){\circle*{25}}
 \put (140,120){\circle*{25}}
 \put (70,160){\circle*{25}}
 \put (0,120){\circle*{25}}

%34
% \put (140,40){\line(0,1){80}} 
 \put (140,50){\circle*{1}}
 \put (140,60){\circle*{1}}
 \put (140,70){\circle*{1}}  
 \put (140,80){\circle*{1}}
 \put (140,90){\circle*{1}}
 \put (140,100){\circle*{1}}
 \put (140,110){\circle*{1}}
 
%12 
 \put (7,36){\circle*{1}}
 \put (14,32){\circle*{1}}
 \put (21,28){\circle*{1}}
 \put (28,24){\circle*{1}}    
 \put (35,20){\circle*{1}}
 \put (42,16){\circle*{1}}  
 \put (49,12){\circle*{1}}
 \put (56,8){\circle*{1}}
 \put (63,4){\circle*{1}}

 %56
 \put (7,124){\circle*{1}}
 \put (14,128){\circle*{1}}
 \put (21,132){\circle*{1}}
 \put (28,136){\circle*{1}}    
 \put (35,140){\circle*{1}}
 \put (42,144){\circle*{1}}  
 \put (49,148){\circle*{1}}
 \put (56,152){\circle*{1}}
 \put (63,156){\circle*{1}}

\end{picture}} +\ket{%
%           5\
%       6       4
%       |
%       1       3
%           2 /     
%
\setlength{\unitlength}{3947sp}
\begin{picture}(160,150)(-10,30)
 \put (0,40){\circle*{25}}
 \put (70,0){\circle*{25}}
 \put (140,40){\circle*{25}}
 \put (140,120){\circle*{25}}
 \put (70,160){\circle*{25}}
 \put (0,120){\circle*{25}}
 
 %\put (0,40){\line(0,1){80}}
  \put (0,60){\circle*{1}} 
  \put (0,70){\circle*{1}}
  \put (0,80){\circle*{1}}
  \put (0,90){\circle*{1}}  
  \put (0,100){\circle*{1}}
  \put (0,110){\circle*{1}}

 \put (133,36){\circle*{1}}
 \put (126,32){\circle*{1}}
 \put (119,28){\circle*{1}}
 \put (112,24){\circle*{1}}    
 \put (105,20){\circle*{1}}
 \put (98,16){\circle*{1}}  
 \put (91,12){\circle*{1}}
 \put (84,8){\circle*{1}}
 \put (77,4){\circle*{1}}

 \put (133,124){\circle*{1}}
 \put (126,128){\circle*{1}}
 \put (119,132){\circle*{1}}
 \put (112,136){\circle*{1}}    
 \put (105,140){\circle*{1}}
 \put (98,144){\circle*{1}}  
 \put (91,148){\circle*{1}}
 \put (84,152){\circle*{1}}
 \put (77,156){\circle*{1}}   
\end{picture}} \bra{%
%           5\
%       6       4
%       |
%       1       3
%           2 /     
%
\setlength{\unitlength}{3947sp}
\begin{picture}(160,150)(-10,30)
 \put (0,40){\circle*{25}}
 \put (70,0){\circle*{25}}
 \put (140,40){\circle*{25}}
 \put (140,120){\circle*{25}}
 \put (70,160){\circle*{25}}
 \put (0,120){\circle*{25}}
 
 %\put (0,40){\line(0,1){80}}
  \put (0,60){\circle*{1}} 
  \put (0,70){\circle*{1}}
  \put (0,80){\circle*{1}}
  \put (0,90){\circle*{1}}  
  \put (0,100){\circle*{1}}
  \put (0,110){\circle*{1}}

 \put (133,36){\circle*{1}}
 \put (126,32){\circle*{1}}
 \put (119,28){\circle*{1}}
 \put (112,24){\circle*{1}}    
 \put (105,20){\circle*{1}}
 \put (98,16){\circle*{1}}  
 \put (91,12){\circle*{1}}
 \put (84,8){\circle*{1}}
 \put (77,4){\circle*{1}}

 \put (133,124){\circle*{1}}
 \put (126,128){\circle*{1}}
 \put (119,132){\circle*{1}}
 \put (112,136){\circle*{1}}    
 \put (105,140){\circle*{1}}
 \put (98,144){\circle*{1}}  
 \put (91,148){\circle*{1}}
 \put (84,152){\circle*{1}}
 \put (77,156){\circle*{1}}   
\end{picture}}\right)\nn\\
&&~~~~-t'\left(\ket{%
%           5\
%       6       4
%       
%       1------ 3
%           2     
%
\setlength{\unitlength}{3947sp}
\begin{picture}(160,150)(-10,30)
 \put (0,40){\circle*{25}}
 \put (70,0){\circle*{25}}
 \put (140,40){\circle*{25}}
 \put (140,120){\circle*{25}}
 \put (70,160){\circle*{25}}
 \put (0,120){\circle*{25}}
 
 %13
 %\put (0,40){\line(1,0){140}}
  \put (10,40){\circle*{1}}
  \put (20,40){\circle*{1}}
  \put (30,40){\circle*{1}}
  \put (40,40){\circle*{1}}
  \put (50,40){\circle*{1}}
  \put (60,40){\circle*{1}}
  \put (70,40){\circle*{1}}
  \put (80,40){\circle*{1}}
  \put (90,40){\circle*{1}}
  \put (100,40){\circle*{1}}
  \put (110,40){\circle*{1}}
  \put (120,40){\circle*{1}} 
  \put (130,40){\circle*{1}}

%45
 \put (133,124){\circle*{1}}
 \put (126,128){\circle*{1}}
 \put (119,132){\circle*{1}}
 \put (112,136){\circle*{1}}    
 \put (105,140){\circle*{1}}
 \put (98,144){\circle*{1}}  
 \put (91,148){\circle*{1}}
 \put (84,152){\circle*{1}}
 \put (77,156){\circle*{1}}   
\end{picture}} \bra{%
%           5\
%       6       4
%       
%       1------ 3
%           2     
%
\setlength{\unitlength}{3947sp}
\begin{picture}(160,150)(-10,30)
 \put (0,40){\circle*{25}}
 \put (70,0){\circle*{25}}
 \put (140,40){\circle*{25}}
 \put (140,120){\circle*{25}}
 \put (70,160){\circle*{25}}
 \put (0,120){\circle*{25}}
 
%15
 \put (7,52){\circle*{1}}
 \put (14,64){\circle*{1}}
 \put (21,76){\circle*{1}}
 \put (28,88){\circle*{1}}    
 \put (35,100){\circle*{1}}
 \put (42,112){\circle*{1}}  
 \put (49,124){\circle*{1}}
 \put (56,136){\circle*{1}}
 \put (63,148){\circle*{1}} 
 
%34
 %\put (140,40){\line(0,1){80}}
 \put (140,50){\circle*{1}}
 \put (140,60){\circle*{1}}
 \put (140,70){\circle*{1}}
 \put (140,80){\circle*{1}}
 \put (140,90){\circle*{1}}
 \put (140,100){\circle*{1}} 
 \put (140,110){\circle*{1}}
   
\end{picture}} +h.c.\right)+V'\left(\ket{%
%           5\
%       6       4
%       
%       1------ 3
%           2     
%
\setlength{\unitlength}{3947sp}
\begin{picture}(160,150)(-10,30)
 \put (0,40){\circle*{25}}
 \put (70,0){\circle*{25}}
 \put (140,40){\circle*{25}}
 \put (140,120){\circle*{25}}
 \put (70,160){\circle*{25}}
 \put (0,120){\circle*{25}}
 
 %13
 %\put (0,40){\line(1,0){140}}
  \put (10,40){\circle*{1}}
  \put (20,40){\circle*{1}}
  \put (30,40){\circle*{1}}
  \put (40,40){\circle*{1}}
  \put (50,40){\circle*{1}}
  \put (60,40){\circle*{1}}
  \put (70,40){\circle*{1}}
  \put (80,40){\circle*{1}}
  \put (90,40){\circle*{1}}
  \put (100,40){\circle*{1}}
  \put (110,40){\circle*{1}}
  \put (120,40){\circle*{1}} 
  \put (130,40){\circle*{1}}

%45
 \put (133,124){\circle*{1}}
 \put (126,128){\circle*{1}}
 \put (119,132){\circle*{1}}
 \put (112,136){\circle*{1}}    
 \put (105,140){\circle*{1}}
 \put (98,144){\circle*{1}}  
 \put (91,148){\circle*{1}}
 \put (84,152){\circle*{1}}
 \put (77,156){\circle*{1}}   
\end{picture}}\bra{%
%           5\
%       6       4
%       
%       1------ 3
%           2     
%
\setlength{\unitlength}{3947sp}
\begin{picture}(160,150)(-10,30)
 \put (0,40){\circle*{25}}
 \put (70,0){\circle*{25}}
 \put (140,40){\circle*{25}}
 \put (140,120){\circle*{25}}
 \put (70,160){\circle*{25}}
 \put (0,120){\circle*{25}}
 
 %13
 %\put (0,40){\line(1,0){140}}
  \put (10,40){\circle*{1}}
  \put (20,40){\circle*{1}}
  \put (30,40){\circle*{1}}
  \put (40,40){\circle*{1}}
  \put (50,40){\circle*{1}}
  \put (60,40){\circle*{1}}
  \put (70,40){\circle*{1}}
  \put (80,40){\circle*{1}}
  \put (90,40){\circle*{1}}
  \put (100,40){\circle*{1}}
  \put (110,40){\circle*{1}}
  \put (120,40){\circle*{1}} 
  \put (130,40){\circle*{1}}

%45
 \put (133,124){\circle*{1}}
 \put (126,128){\circle*{1}}
 \put (119,132){\circle*{1}}
 \put (112,136){\circle*{1}}    
 \put (105,140){\circle*{1}}
 \put (98,144){\circle*{1}}  
 \put (91,148){\circle*{1}}
 \put (84,152){\circle*{1}}
 \put (77,156){\circle*{1}}   
\end{picture}} +\ket{%
%           5\
%       6       4
%       
%       1------ 3
%           2     
%
\setlength{\unitlength}{3947sp}
\begin{picture}(160,150)(-10,30)
 \put (0,40){\circle*{25}}
 \put (70,0){\circle*{25}}
 \put (140,40){\circle*{25}}
 \put (140,120){\circle*{25}}
 \put (70,160){\circle*{25}}
 \put (0,120){\circle*{25}}
 
%15
 \put (7,52){\circle*{1}}
 \put (14,64){\circle*{1}}
 \put (21,76){\circle*{1}}
 \put (28,88){\circle*{1}}    
 \put (35,100){\circle*{1}}
 \put (42,112){\circle*{1}}  
 \put (49,124){\circle*{1}}
 \put (56,136){\circle*{1}}
 \put (63,148){\circle*{1}} 
 
%34
 %\put (140,40){\line(0,1){80}}
 \put (140,50){\circle*{1}}
 \put (140,60){\circle*{1}}
 \put (140,70){\circle*{1}}
 \put (140,80){\circle*{1}}
 \put (140,90){\circle*{1}}
 \put (140,100){\circle*{1}} 
 \put (140,110){\circle*{1}}
   
\end{picture}}\bra{%
%           5\
%       6       4
%       
%       1------ 3
%           2     
%
\setlength{\unitlength}{3947sp}
\begin{picture}(160,150)(-10,30)
 \put (0,40){\circle*{25}}
 \put (70,0){\circle*{25}}
 \put (140,40){\circle*{25}}
 \put (140,120){\circle*{25}}
 \put (70,160){\circle*{25}}
 \put (0,120){\circle*{25}}
 
%15
 \put (7,52){\circle*{1}}
 \put (14,64){\circle*{1}}
 \put (21,76){\circle*{1}}
 \put (28,88){\circle*{1}}    
 \put (35,100){\circle*{1}}
 \put (42,112){\circle*{1}}  
 \put (49,124){\circle*{1}}
 \put (56,136){\circle*{1}}
 \put (63,148){\circle*{1}} 
 
%34
 %\put (140,40){\line(0,1){80}}
 \put (140,50){\circle*{1}}
 \put (140,60){\circle*{1}}
 \put (140,70){\circle*{1}}
 \put (140,80){\circle*{1}}
 \put (140,90){\circle*{1}}
 \put (140,100){\circle*{1}} 
 \put (140,110){\circle*{1}}
   
\end{picture}}\right) \nn\\
&&\!\!-t^{\prime\prime}\left(\ket{%
%           5\
%       6       4
%       
%       1------ 3
%           2     
%
\setlength{\unitlength}{3947sp}
\begin{picture}(160,150)(-10,30)
 \put (0,40){\circle*{25}}
 \put (70,0){\circle*{25}}
 \put (140,40){\circle*{25}}
 \put (140,120){\circle*{25}}
 \put (70,160){\circle*{25}}
 \put (0,120){\circle*{25}}
 
%13
  %\put (0,40){\line(1,0){140}}  
  \put (10,40){\circle*{1}}
  \put (20,40){\circle*{1}}
  \put (30,40){\circle*{1}}
  \put (40,40){\circle*{1}}
  \put (50,40){\circle*{1}}
  \put (60,40){\circle*{1}}
  \put (70,40){\circle*{1}}
  \put (80,40){\circle*{1}}
  \put (90,40){\circle*{1}}
  \put (100,40){\circle*{1}}
  \put (110,40){\circle*{1}}
  \put (120,40){\circle*{1}} 
  \put (130,40){\circle*{1}}
 
%46
 %\put (0,120){\line(1,0){140}}   
  \put (10,120){\circle*{1}}
  \put (20,120){\circle*{1}}
  \put (30,120){\circle*{1}}
  \put (40,120){\circle*{1}}
  \put (50,120){\circle*{1}}
  \put (60,120){\circle*{1}}
  \put (70,120){\circle*{1}}
  \put (80,120){\circle*{1}}
  \put (90,120){\circle*{1}}
  \put (100,120){\circle*{1}}
  \put (110,120){\circle*{1}}
  \put (120,120){\circle*{1}} 
  \put (130,120){\circle*{1}}
\end{picture}} \bra{%
%           5\
%       6       4
%       
%       1------ 3
%           2     
%
\setlength{\unitlength}{3947sp}
\begin{picture}(160,150)(-10,30)
 \put (0,40){\circle*{25}}
 \put (70,0){\circle*{25}}
 \put (140,40){\circle*{25}}
 \put (140,120){\circle*{25}}
 \put (70,160){\circle*{25}}
 \put (0,120){\circle*{25}}

%16
 %\put (0,40){\line(0,1){80}}
  \put (0,50){\circle*{1}}
  \put (0,60){\circle*{1}}
  \put (0,70){\circle*{1}}
  \put (0,80){\circle*{1}}
  \put (0,90){\circle*{1}}
  \put (0,100){\circle*{1}}
  \put (0,110){\circle*{1}}
 
 %34
   %\put (140,40){\line(0,1){80}}
  \put (140,50){\circle*{1}}
  \put (140,60){\circle*{1}}
  \put (140,70){\circle*{1}}
  \put (140,80){\circle*{1}}
  \put (140,90){\circle*{1}}
  \put (140,100){\circle*{1}}
  \put (140,110){\circle*{1}}

\end{picture}} +h.c.\right)+\lambda V^{\prime\prime}\ket{%
%           5\
%       6       4
%       
%       1------ 3
%           2     
%
\setlength{\unitlength}{3947sp}
\begin{picture}(160,150)(-10,30)
 \put (0,40){\circle*{25}}
 \put (70,0){\circle*{25}}
 \put (140,40){\circle*{25}}
 \put (140,120){\circle*{25}}
 \put (70,160){\circle*{25}}
 \put (0,120){\circle*{25}}

%16
 %\put (0,40){\line(0,1){80}}
  \put (0,50){\circle*{1}}
  \put (0,60){\circle*{1}}
  \put (0,70){\circle*{1}}
  \put (0,80){\circle*{1}}
  \put (0,90){\circle*{1}}
  \put (0,100){\circle*{1}}
  \put (0,110){\circle*{1}}
 
 %34
   %\put (140,40){\line(0,1){80}}
  \put (140,50){\circle*{1}}
  \put (140,60){\circle*{1}}
  \put (140,70){\circle*{1}}
  \put (140,80){\circle*{1}}
  \put (140,90){\circle*{1}}
  \put (140,100){\circle*{1}}
  \put (140,110){\circle*{1}}

\end{picture}}\bra{%
%           5\
%       6       4
%       
%       1------ 3
%           2     
%
\setlength{\unitlength}{3947sp}
\begin{picture}(160,150)(-10,30)
 \put (0,40){\circle*{25}}
 \put (70,0){\circle*{25}}
 \put (140,40){\circle*{25}}
 \put (140,120){\circle*{25}}
 \put (70,160){\circle*{25}}
 \put (0,120){\circle*{25}}

%16
 %\put (0,40){\line(0,1){80}}
  \put (0,50){\circle*{1}}
  \put (0,60){\circle*{1}}
  \put (0,70){\circle*{1}}
  \put (0,80){\circle*{1}}
  \put (0,90){\circle*{1}}
  \put (0,100){\circle*{1}}
  \put (0,110){\circle*{1}}
 
 %34
   %\put (140,40){\line(0,1){80}}
  \put (140,50){\circle*{1}}
  \put (140,60){\circle*{1}}
  \put (140,70){\circle*{1}}
  \put (140,80){\circle*{1}}
  \put (140,90){\circle*{1}}
  \put (140,100){\circle*{1}}
  \put (140,110){\circle*{1}}

\end{picture}} +\frac{V^{\prime\prime}}{\lambda}\ket{%
%           5\
%       6       4
%       
%       1------ 3
%           2     
%
\setlength{\unitlength}{3947sp}
\begin{picture}(160,150)(-10,30)
 \put (0,40){\circle*{25}}
 \put (70,0){\circle*{25}}
 \put (140,40){\circle*{25}}
 \put (140,120){\circle*{25}}
 \put (70,160){\circle*{25}}
 \put (0,120){\circle*{25}}
 
%13
  %\put (0,40){\line(1,0){140}}  
  \put (10,40){\circle*{1}}
  \put (20,40){\circle*{1}}
  \put (30,40){\circle*{1}}
  \put (40,40){\circle*{1}}
  \put (50,40){\circle*{1}}
  \put (60,40){\circle*{1}}
  \put (70,40){\circle*{1}}
  \put (80,40){\circle*{1}}
  \put (90,40){\circle*{1}}
  \put (100,40){\circle*{1}}
  \put (110,40){\circle*{1}}
  \put (120,40){\circle*{1}} 
  \put (130,40){\circle*{1}}
 
%46
 %\put (0,120){\line(1,0){140}}   
  \put (10,120){\circle*{1}}
  \put (20,120){\circle*{1}}
  \put (30,120){\circle*{1}}
  \put (40,120){\circle*{1}}
  \put (50,120){\circle*{1}}
  \put (60,120){\circle*{1}}
  \put (70,120){\circle*{1}}
  \put (80,120){\circle*{1}}
  \put (90,120){\circle*{1}}
  \put (100,120){\circle*{1}}
  \put (110,120){\circle*{1}}
  \put (120,120){\circle*{1}} 
  \put (130,120){\circle*{1}}
\end{picture}} \bra{%
%           5\
%       6       4
%       
%       1------ 3
%           2     
%
\setlength{\unitlength}{3947sp}
\begin{picture}(160,150)(-10,30)
 \put (0,40){\circle*{25}}
 \put (70,0){\circle*{25}}
 \put (140,40){\circle*{25}}
 \put (140,120){\circle*{25}}
 \put (70,160){\circle*{25}}
 \put (0,120){\circle*{25}}
 
%13
  %\put (0,40){\line(1,0){140}}  
  \put (10,40){\circle*{1}}
  \put (20,40){\circle*{1}}
  \put (30,40){\circle*{1}}
  \put (40,40){\circle*{1}}
  \put (50,40){\circle*{1}}
  \put (60,40){\circle*{1}}
  \put (70,40){\circle*{1}}
  \put (80,40){\circle*{1}}
  \put (90,40){\circle*{1}}
  \put (100,40){\circle*{1}}
  \put (110,40){\circle*{1}}
  \put (120,40){\circle*{1}} 
  \put (130,40){\circle*{1}}
 
%46
 %\put (0,120){\line(1,0){140}}   
  \put (10,120){\circle*{1}}
  \put (20,120){\circle*{1}}
  \put (30,120){\circle*{1}}
  \put (40,120){\circle*{1}}
  \put (50,120){\circle*{1}}
  \put (60,120){\circle*{1}}
  \put (70,120){\circle*{1}}
  \put (80,120){\circle*{1}}
  \put (90,120){\circle*{1}}
  \put (100,120){\circle*{1}}
  \put (110,120){\circle*{1}}
  \put (120,120){\circle*{1}} 
  \put (130,120){\circle*{1}}
\end{picture}}\Big],
\eea
where various parameters have similar meaning as in the case of square lattice, and it is implicit that all terms are to be summed over symmetry related orientations.

In defining these models, we
have restricted the Hilbert space
by excluding configurations with crossed dimers.
We do this for two reasons:
Firstly,
more microscopically,
the dimers are thought to be representations of
spin singlets (or valence bonds)
which are assumed to be the building blocks of the low energy subspace of an underlying quantum spin 1/2 problem with strong frustration.  However,
there are only two linearly independent singlet states corresponding to four spin 1/2's on a plaquette, or in other words, if we identify dimer configurations with orthogonalized versions of valence bond states, then
$\ket{% 
%       2         4        6
%       -         -
%       -         -
%       -         -
%       1         3        5
%
\setlength{\unitlength}{3947sp}
\begin{picture}(160,120)(-20,10)
 \put (0,0){\circle*{25}}
 \put (120,0){\circle*{25}}
 \put (0,120){\circle*{25}}
 \put (120,120){\circle*{25}}
 
 %\put (0,0){\line(0,1){120}} 
 %\put (120,0){\line(0,1){120}}   
 \put (10,10){\circle*{1}}
 \put (20,20){\circle*{1}}
 \put (30,30){\circle*{1}}
 \put (40,40){\circle*{1}}    
 \put (50,50){\circle*{1}}
 \put (60,60){\circle*{1}}  
 \put (70,70){\circle*{1}}
 \put (80,80){\circle*{1}}    
 \put (90,90){\circle*{1}}
 \put (100,100){\circle*{1}}    
 \put (110,110){\circle*{1}}
 
 \put (10,110){\circle*{1}}
 \put (20,100){\circle*{1}}
 \put (30,90){\circle*{1}}
 \put (40,80){\circle*{1}}    
 \put (50,70){\circle*{1}}
 \put (60,60){\circle*{1}}  
 \put (70,50){\circle*{1}}
 \put (80,40){\circle*{1}}    
 \put (90,30){\circle*{1}}
 \put (100,20){\circle*{1}}    
 \put (110,10){\circle*{1}} 
\end{picture}}\propto -(\ket{% 
%       2         4        6
%       -         -
%       -         -
%       -         -
%       1         3        5
%
\setlength{\unitlength}{3947sp}
\begin{picture}(160,120)(-20,10)
 \put (0,0){\circle*{25}}
 \put (120,0){\circle*{25}}
 \put (0,120){\circle*{25}}
 \put (120,120){\circle*{25}}
 
 %\put (0,0){\line(0,1){120}} 
 %\put (120,0){\line(0,1){120}}   
 \put (0,10){\circle*{1}}
 \put (0,20){\circle*{1}}
 \put (0,30){\circle*{1}}
 \put (0,40){\circle*{1}}    
 \put (0,50){\circle*{1}}
 \put (0,60){\circle*{1}}  
 \put (0,70){\circle*{1}}
 \put (0,80){\circle*{1}}    
 \put (0,90){\circle*{1}}
 \put (0,100){\circle*{1}}    
 \put (0,110){\circle*{1}}
 
 \put (120,10){\circle*{1}}
 \put (120,20){\circle*{1}}
 \put (120,30){\circle*{1}}
 \put (120,40){\circle*{1}}    
 \put (120,50){\circle*{1}}
 \put (120,60){\circle*{1}}  
 \put (120,70){\circle*{1}}
 \put (120,80){\circle*{1}}    
 \put (120,90){\circle*{1}}
 \put (120,100){\circle*{1}}    
 \put (120,110){\circle*{1}} 
\end{picture}} +\ket{% 
%       2--------4        6
%
%          
%
%       1--------3        5
%
\setlength{\unitlength}{3947sp}
\begin{picture}(160,120)(-20,10)
 \put (0,0){\circle*{25}}
 \put (120,0){\circle*{25}}
 \put (0,120){\circle*{25}}
 \put (120,120){\circle*{25}}
 %\put (0,120){\line(1,0){120}} 
 %\put (0,0){\line(1,0){120}}  

 \put (10,0){\circle*{1}}
 \put (20,0){\circle*{1}}
 \put (30,0){\circle*{1}}
 \put (40,0){\circle*{1}}    
 \put (50,0){\circle*{1}}
 \put (60,0){\circle*{1}}  
 \put (70,0){\circle*{1}}
 \put (80,0){\circle*{1}}    
 \put (90,0){\circle*{1}}
 \put (100,0){\circle*{1}}    
 \put (110,0){\circle*{1}}
 
 \put (10,120){\circle*{1}}
 \put (20,120){\circle*{1}}
 \put (30,120){\circle*{1}}
 \put (40,120){\circle*{1}}    
 \put (50,120){\circle*{1}}
 \put (60,120){\circle*{1}}  
 \put (70,120){\circle*{1}}
 \put (80,120){\circle*{1}}    
 \put (90,120){\circle*{1}}
 \put (100,120){\circle*{1}}    
 \put (110,120){\circle*{1}} 
    
\end{picture}})$.
(It is an open question whether the remaining states with first and non-crossing second neighbor valence bonds are linearly independent\cite{seidel}.)
Secondly, for the model as defined, crossed dimers would be non-dynamical unless we were to include additional terms in the Hamiltonian.

For the system with a torus or cylinder geometry, it is known that dimer configurations
can be classified according to topological labels or winding numbers;
dimer configurations with different winding numbers cannot be connected by local moves of dimers. On the square and honeycomb lattices, configurations with only first neighbor dimers are characterized by integer winding numbers\cite{rk} due to the bipartiteness of the model. However, for the dimer configurations allowing both first and second neighbor dimers, the bipartiteness is lost and on a 2-torus there are only four topological sectors labeled by $W=(W_x,W_y)$, where $W_i$ is odd or even. A simple way to determine $W_x$ is to draw a vertical line across the whole system and see how many dimers it cuts whose parity is then $W_x$, and $W_y$ is defined similarly in terms of the even or oddness of the number of dimers crossing  a horizontal line.
More generally,  the number and character of the topological sectors depends on the boundary conditions in ways that are straightforward to determine.

\begin{figure}[b]
\subfigure[]{
\includegraphics[scale=0.32]{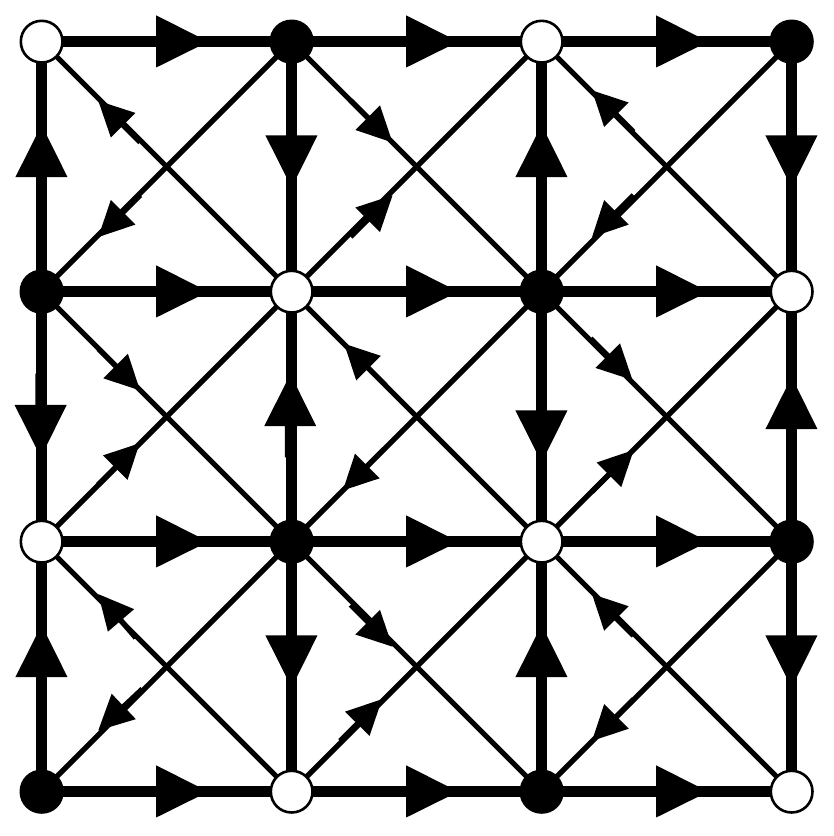}}~~~
\subfigure[]{
\includegraphics[scale=0.32]{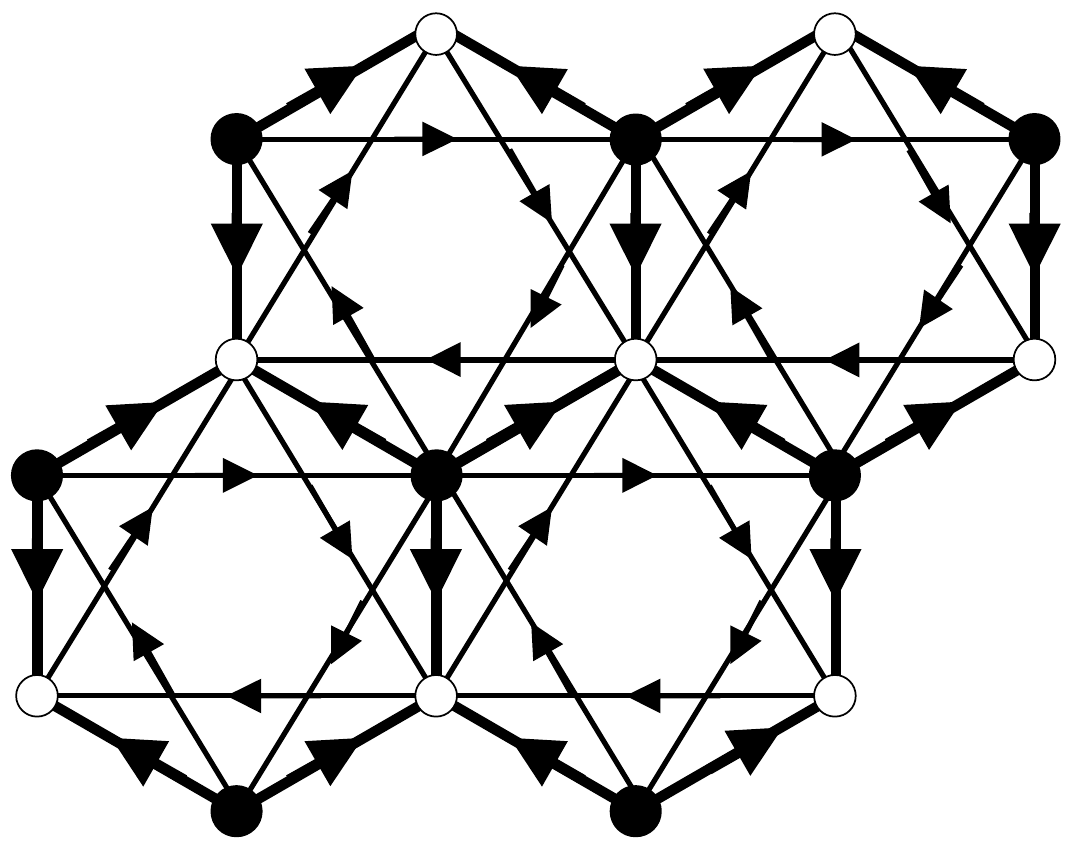}}
\caption{%The schematic representation of the square lattice and honeycomb lattice.
 The arrow pattern on %both lattices is shown explicitly.
 on the square and honeycomb lattices used in defining the Grassmann path integral representation of the classical dimer model.
 The black and white sites are the two sublattices. }
\label{fig:latt}
\end{figure}

{\bf The RK point:} For generic parameters
the exact ground state
is not known.  However, at the
so-called Rohksar-Kivelson (RK) points, exact ground states of both the models can be
determined explicitly.  For both Hamiltonians [\Eq{eq:square} and \Eq{eq:honeycomb}], the generalized RK points are given by
\bea
t=V, ~t'=V', ~\textrm{and}~%\tau=U.
t^{\prime\prime}=V^{\prime\prime},
\eea
and any finite $\lambda$.
At these RK points, both \Eq{eq:square} and \Eq{eq:honeycomb}
can be expressed as a sum of projection operators and
can thus be shown to be positive semi-definite\cite{rk}, with groundstates:
\bea\label{eq:rk}
\ket{\Psi_0}=\sum_{c} \lambda^{n_s(c)/2} \ket{c},
\eea
where the summation is over all dimer configurations, $c$, %\ket{c}$,
 in a given topological sector and $n_s(c)$ is the number of second-neighbor dimers in $c$.    That this is an exact ground-state can be seen
 by explicitly checking that it is anihilated by the Hamiltonian [\Eq{eq:square} or \Eq{eq:honeycomb}].
On the torus, there are four
sectors so the ground states have fourfold topological degeneracy\cite{krs,wen}.

It remains to characterize the phases described by these exact ground-states in terms of the behavior of the dimer-dimer correlation functions.  For the RK wave function with only first neighbor dimers,
the ground state
is known to correspond to a quantum multi-crtical point
with power-law decay of
correlation functions.  However, as we shall show, the
wave function described by \Eq{eq:rk} is
more akin to that of the dimer model on the triangular lattice in the sense that
all groundstate correlation functions fall exponentially with distance\cite{moessner}.

{\bf Path integral representation:} The dimer-dimer correlation function in a ground state RK wave function $\ket{\Psi_0}$ is defined as
\bea
\avg{D_{ij}D_{i'j'}}\equiv\frac{\bra{\Psi_0} D_{ij}D_{i'j'}\ket{\Psi_0}}{\braket{\Psi_0}{\Psi_0}},
\eea
where $D_{ij}$ denote dimer operators on the link $(ij)$ which is 1 when the link is occupied by a dimer and 0 otherwise.
This is equivalent to the correlations of a classical dimer model in which
the wave function normalization,  $Z=\braket{\Psi_0}{\Psi_0}$, plays the role of the partition function. For
wave functions involving only first neighbor dimers, it is known that dimer densities and dimer-dimer correlation functions can be computed
in terms of a path integral representation of non-interacting Grassmann fields, since a consistent arrow pattern determining the action of Grassmann fields is always possible for a planar graph\cite{pfaffian}.
Due to the appearance of second neighbor dimers, the graph is no longer planar and the Pfaffian method does not work directly. Nonetheless, the partition function can still be {\it exactly} expressed by the following path integral representation of {\it interacting} Grassmann fields (see the proof in Appendix):
\bea
Z&=&\int [da] e^{-S},\label{eq:pathint}\\
S&=&\sum_{ij} it_{ij}a_i a_j + \sum_{\avg{ijkl}} (it_{ij}a_i a_j)(it_{kl}a_k a_l),\label{eq:action}
\eea
where $a_i$ are Grassmann numbers and $[da]=\prod_{i=1}^N da_i$ ($N$ is the number of sites), $t_{ij}=\pm 1$ on the first neighbor links,
$t_{ij}=\pm \lambda$ on second neighbor links, and $\avg{ijkl}$ denotes a pair of crossing second neighbor links $ij$ and $kl$.
Here, the sign of $t_{ij}$ is determined from the pattern of arrows shown in  \Fig{fig:latt}, such that
$t_{ij}$ is positive if the
arrow on the link $(ij)$ points
 from $i$ to $j$, and negative if it points from $j$ to $i$.
 (The choice of a particular pattern of arrows amounts to a choice of gauge.)

Since the new thing here is the four-field ``interaction'' term, it
worth discussing its origin heuristically.
In the absence of the interaction term,
the path integral has a contribution from dimer configurations with crossed dimers with an associated amplitude factor,
$[-t_{ij} t_{kl}]$, which can be negative.
 The interaction term likewise
 generates
 crossed dimers with an amplitude
 $t_{ij}t_{kl}$, which is just what is needed to cancel the spurious term in the absence of interactions.
 The proof is in Appendix.

\begin{figure}[b]
\includegraphics[scale=0.35]{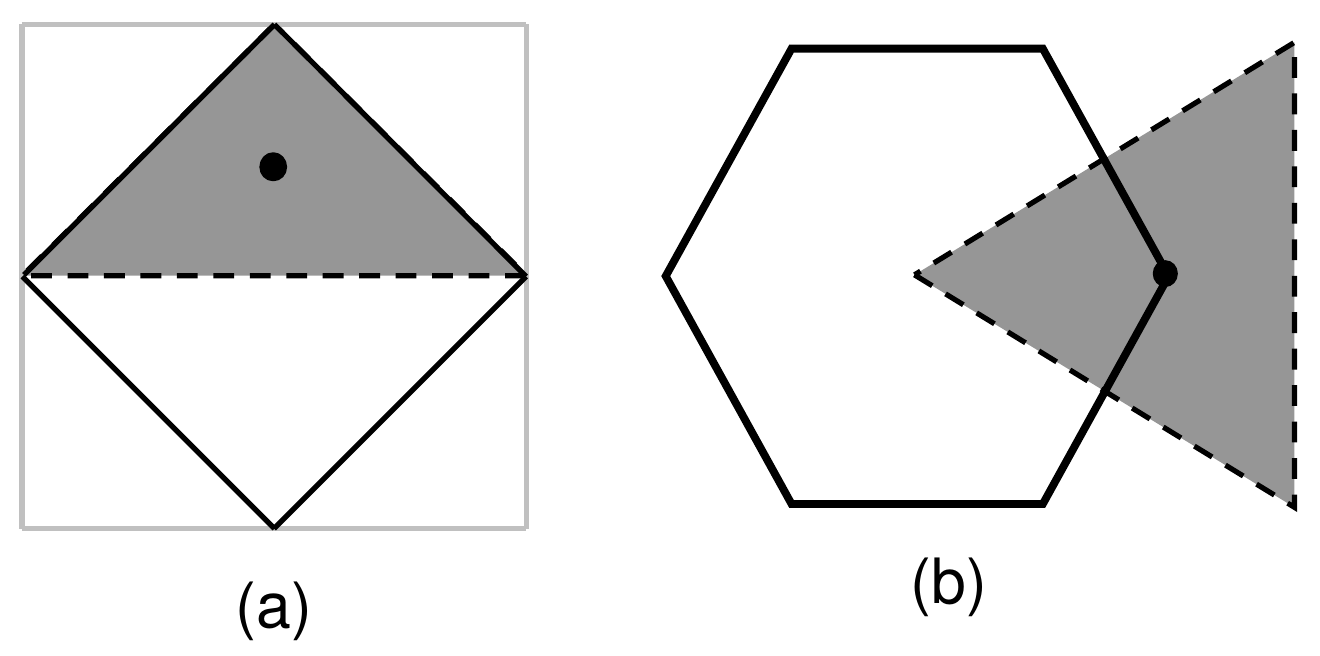}
\caption{The Brillouin zone (BZ) for the square and honeycomb lattices:  The solid lines mark the boundaries of the full BZ corresponding to the arrow patterns in Fig. 1, with two sites per unit cell.  The shaded area is the ``half BZ'' over which the $\vec k$-sum in Eq. 9 is carried out.  The %(red)
dot indicates the location of the Dirac node, $\vec K$, in the limit $\lambda\to 0$. }
\label{fig:half}
\end{figure}

{\bf Massive Thirring model:} The path integral in  \Eq{eq:pathint} describes the classical dynamics of fermionic fields in 2 spatial dimensions which can be mapped to a 1+1 dimensional quantum theory by treating one spatial dimension, say $y$, as time $t$.

For the square lattice, the arrow pattern we have chosen doubles the unit cell; for the honeycomb lattice, the unit cell already contains two sites, and no further increase is necessary in order to define a consistent pattern of arrows;
we will thus define a new two-component field with a pseudo-spin index, $\alpha$,  which labels the sublattice.  Translation symmetry allows us to block diagonalize the non-interacting part of the Hamiltonian by Fourier transform in terms of a new set of {\it complex} Grassmann fields
\bea
b_{\vec k,\alpha }=\sqrt{\frac1{N}}\sum_{j\in \alpha}e^{i\vec k\cdot \vec r_j} a_{j},
\eea
where  $b^\dagger_{\vec k,\alpha} = b_{-\vec k,\alpha}$,
 since $a_j$ are real Grassmann fields.
The non-interacting part of the action
is then
\bea
S_0=\sum_{\vec k\in \frac12\textrm{BZ}} b^\dag_{\vec k} h_{\vec k} b_{\vec k},
\label{S0}
\eea
where
the sum runs over only half the Brillouin zone  (See Fig. 2.) (since we have combined real fields with $\vec k$ and $-\vec k$ into a single complex field), and
\bea
h_{\vec k}=2\left[\ba{cc}
2\lambda \cos k_x\sin k_y & -\sin k_x + i\cos k_y\\
-\sin k_x - i\cos k_y& -2\lambda \cos k_x\sin k_y
\ea\right]
\eea
and
\bea
h_{\vec k}=\sum_{j=1}^3\left[\ba{cc}
2\lambda \sin[\vec k \cdot (\hat e_{j+1}-\hat e_j)] &  ie^{i\vec k\cdot \hat e_j}\\
 -ie^{-i\vec k\cdot \hat e_j} & -2\lambda \sin[\vec k \cdot (\hat e_{j+1}-\hat e_j)]
\ea\right],
\nonumber
\eea
for the square and honeycomb lattice, respectively. Here $\hat e_j$ are the vectors  connecting first neighbors of the honeycomb lattice: $\hat e_1=(0,-1)$, $\hat e_2 = (\sqrt{3}/2,1/2)$, $\hat e_3 =  (-\sqrt{3}/2,1/2)$, and $\hat e_{j+3}\equiv \hat e_j$.

When $\lambda=0$, the spectrum of $h_{\vec k}$ is gapless and has a single Dirac point at a momentum labeled
$\vec K$ (the
dot points in \Fig{fig:half}) within the half Brillouin zone:  $\vec K=(0,\pi/2)$ for the square and  $\vec K=(4\pi/3\sqrt{3},0)$ for the honeycomb lattice. A finite $\lambda$ opens up a gap in the spectrum and is the mass term of the Dirac fermions.
Indeed, taking the continuum limit of Eq. \ref{S0}
for small $\lambda$
produces a formal equivalence to the 1+1 D Dirac action:
\bea
S_0=v_F\int d x d y \psi^\dag[i\partial_y\sigma^x-i\partial_x \sigma^y + m_0 \sigma^z]\psi,
\eea
where for the square lattice,  $v_F=2$ and $m_0=2\lambda$ while for the honeycomb lattice, $v_F=\frac32$ and  $m_0= 2\sqrt3\lambda$.
Here $\psi_\alpha(\vec r)$ is the slowly varying piece of the inverse Fourier transform of $b_{\vec k,\alpha}$, with the rapidly oscillating factor, $\exp[i\vec K\cdot\vec r]$, removed.
For small $\lambda$,  the interaction term can also be treated in the naive continuum limit, so
\bea\label{eq:2d}
S\!%=
\to S_0  \!-\!g \!\int\! dx dy%\!\left[
%\psi^\dag (i\pa_y\sigma^x\!-\!i\pa_x\sigma^y\!+\!m_0\sigma^z) \psi
 \psi^\dag_1\psi_1\psi^\dag_2\psi_2. %\right].
\eea
where $g=8\lambda^2 $ for the square lattice and $g=24\lambda^2$ for the honeycomb is the strength of attractive interaction.
The
effective interaction is attractive
as it serves to cancel the (negative) weight for crossed dimers generated by $S_0$.

Rescaling the spatial coordinates so that $v_F=1$, identifying $y$ with the imaginary time, $t$, and introducing the notation  $\bar\psi=\psi^\dag\sigma^z$, we see that \Eq{eq:2d} is precisely the Euclidean action of the massive Thirring model.
 The massive  Thirring model is, in turn, exactly solvable by Bethe ansatz\cite{thirring}.  The mass is renormalized as:
\bea
m\approx m_0 e^{-g/v_F\pi}.
\eea
In other words, the mass gap gets renormalized down but survives under weak attractive interactions.
Indeed, the effect of interactions is only perturbative and does not qualitatively change the feature of the non-interacting model, as long as $\lambda$ is small, {\it i.e.}, $ \lambda^2 \log(\frac1{\lambda})\ll 1$.

The finite gap in the spectrum at small but finite $\lambda$ implies that the ground state dimer correlations are short-ranged\cite{paul}. Consequently, the ground states are gapped $Z_2$ quantum spin liquids with a fourfold topological degeneracy on a torus \cite{moessner-sondhi-fradkin}. This is our central result.  %indicating that

{\bf Even-odd effect:} Unambiguously identifying gapped spin-liquid phases is notoriously difficult experimentally since spin-liquid ground states exhibit no conventional broken symmetries.
In this context,
we note that when a
system with a gapped spin-liquid groundstate is placed on a torus with a finite odd-length circumference $L_y$ in one direction, a two-fold degenerate density-wave state ({\it i.e.} which spontaneously breaks translational symmetry) necessarily arises.
However, the density wave order vanishes when $L_y$ is even. This even-odd effect is rooted in the seminal theorem
of Lieb, Schultz and Mattis\cite{lsm,misguich}: For a 1D system with an odd number of electrons per unit cell, there must always exist a distinct state with momentum $\pi$ relative to the ground-state whose energy approaches arbitrarily close the ground state energy in the thermodynamic limit. (This theorem can be applied to the quantum dimer model by associating each dimer with a
singlet pair of electrons \cite{rk}.)  Applying the Lieb-Schultz-Mattis theorem to gapped spin-liquids,
one concludes that this ground-state degeneracy must correspond to translational symmetry breaking for $L_y$ odd.

We have calculated the density-wave order parameters
for  various quantum dimer models on a
torus
of  finite circumference $L_y$. In all cases, the density-wave order vanishes
for even $L_y$. However, for odd and large $L_y$, we find columnar order which decays
as $%\sim\frac{1}{
L_y^{-1/2}\exp[-L_y/2\xi]$, where $\xi$ is the dimer-dimer correlation length in 2D. For the square lattice with small $\lambda$, the columnar density-wave order parameter $C(L_y)$ (defined as the difference between strong and weak bonds) is
\bea
C(L_y)
\approx 2\sqrt{\frac{4\lambda(1-4\lambda^2)} {\pi(1-8\lambda^2)}} \left[\frac{e^{-(L_y+1)/2\xi}}{\sqrt{L_y+1}} +\frac{e^{-(L_y-1)/2\xi}}{\sqrt{L_y-1}}\right],
\nonumber
\eea
for $L_y\gg \xi$,
where $\xi\approx 1/(4\lambda)$.
Both the even-odd effect and exponential decay of the
columnar order with increasing $L_y$ have been observed in DMRG studies of the spin-1/2 $J_1$-$J_2$ Heisenberg model on a square lattice\cite{yao}.

We have obtained similar even-odd results
for the models on the triangular and Kagome lattices.
Because the honeycomb lattice has two sites per unit cell, it is necessary to introduce twisted boundary conditions in order to have a torus with odd $L_y$.  This twist globally lifts the degeneracy between the two, distinct columnar states.
 None-the-less, the magnitude of the bond alternation decays exponentially with increasing $L_y$ in this case, as well.

{\bf Remark:} The present analysis of the ground-state correlations applies only for $\lambda$ small; it is uncertain whether or not there is a transition, possibly to a broken symmetry state, above a nonzero critical $\lambda$.

{\it Acknowledgement}: We would like to thank Paul Fendley, Eduardo Fradkin, Hongchen Jiang, Dung-Hai Lee, Roderich Moessner, Shivaji Sondhi, Wei-Feng Tsai, and Fan Yang for inspiring discussions. This work is partly supported by NSF Grants DMR-0904264 (H.Y.) and DMR-0758356 (S.A.K.).

\section{supplemental material:  proof of the path integral representation}
%\section*{Appendix A. {The proof of the path integral representation}
\renewcommand{\theequation}{A\arabic{equation}}%redefine the command that creates the equation no.
\setcounter{equation}{0}%reset counter
Here, we prove that the path integral  over interacting Grassmann fields produces an exact representation of the Boltzman weight for the classical dimer problem defined by the ground-state expectation-values %values in the ground state
at the generalized RK points on the square and honeycomb lattices.
For simplicity, to avoiding the complication arising from the existence of different topological sectors for the system on a torus, we will first treat the system with open boundary conditions along both direction.  We turn to the problem on a %but the discussion is readily generalized to the torus.
 torus subsequently.
 %We
 Moreover, we first focus on the calculation of the partition function of the classical dimer problem (the wave-function normalization), and turn to the calculation of ground-state correlation functions in the final paragraphs.

%Suppose that $z_n$ is the weighted number of dimer configurations with $n$ crossed pairs of dimers. For $n=0$,
Define $z_0$ to be  the weighted sum of legitimate dimer configurations without crossed dimers which % is also
determines the self-overlap of the ground state RK wave function $\ket{\Psi_0}$: % given by
\bea
z_0=\braket{\Psi_0}{\Psi_0}=\sum_{c} \lambda^{n_s(c)}.
\eea
We further define a set of ``illegitimate'' partition functions, $z_n$, by the same expression, but with the sum over configurations corresponding to all dimer coverings with exactly $n >0$ pairs of crossed dimers.
The partition function from the path integral is $Z=\int [da] e^{-S}$. %We shall
Our goal is to prove that $Z=z_0$. % below.

%First, we have
It is directly clear that
\bea
Z&=&\sum_{m=0}^{N_d/2} Z_m,\nn\\
Z_m&=&\int [da] \frac{(-S)^{N_d-m}}{(N_d-m)!},
\eea
where %$m$ in
$Z_m$ is the $m^{th}$ term in the expansion of $Z$ in powers of the interaction strength ({\it i.e.} the second term in Eq. 7), % , {\it i.e.} order of interaction terms appearing in the expansion of the partition function
 and $N_d$ is the number of dimers or one half the number of lattice sites $N_d=N/2$ (which we assume to be even for simplicity). %Orders higher
 Higher order terms, $Z_m$ with  $m > N_d/2$, vanish identically due to the nature of the Grassmann  integral. For $Z_0$, we obtain
\bea
Z_0&=&\int [da]\frac{(-S)^{N_d}}{N_d!},\nn\\
&=&z_0- z_1+z_2 +\cdots+(-1)^{\frac{N_d}2}z_{N_d/2},\nn\\
&=&\sum_{n=0}^{N_d/2}(-1)^n z_n,
\eea
where the sign $(-1)^n$ is due to the ``incorrect'' arrow pattern involving crossed pairs of dimers generated by the non-interacting term in $S$, and the factor $1/N_d!$ is canceled by the expansion of $(-S)^{N_d}$ in terms of dimer configurations. %To obtain $Z=z_0$, terms $z_{n>0}$ should be canceled by the interaction effects.

The first term involving interactions is
\bea
Z_1&=&\int[da] \frac{(-S)^{N_d-1}}{(N_d-1)!}\nn\\&=&z_1-2z_2+\cdots -(-1)^{\frac{N_d}2}{N_d/2\choose 1}z_{N_d/2}\nn\\
&=&-\sum_{n=1}^{N_d/2}(-1)^n{n \choose 1}z_n,
\eea
where the coefficients ${n\choose 1}$ above is due to the fact that there are ${n\choose 1}$ number of ways to choose a crossed plaquette generated by the interaction term. In general, we obtain
\bea
Z_m&=&\int[da]  \frac{(-S)^{N_d-m}}{(N_d-m)!},\nn\\
&=&(-1)^m\sum_{n=m}^{N_d/2} (-1)^n {n\choose m}z_n.
\eea
By denoting $Z=\sum_{n=0}^{N_d/2} f_n z_n$, we finally obtain
\bea
f_n&=&\sum_{m=0}^n (-1)^n {n\choose m}(-1)^m,\nn\\
&=&(-1)^n (1-1)^n,\nn\\
&=&\left\{
\ba{cc}
1, & n=0,\\
0, & n>0.
\ea\right.
\eea
In other words, we have proved that $Z=z_0$.

The path integral representation above is exact for a system with open boundary conditions along both directions. For a system on torus, %it has a slight complication from
the dimer configurations break into topological sectors, $W=(W_x,W_y)$ where $W_a$  can take on values $e$ or $o$ depending on whether the number of dimers crossing a reference line in the direction perpendicular to $\hat e_a$ is even or odd. The RK wave function in a given topological sector
is given by a combination of path integrals of Grassmann fields with different boundary condition.
Let $z_W$ denote the RK wave function self-overlap in sector $W$.
For $W=(e,e)$,  $z_{(e,e)}=(Z_{pp}+Z_{ap}+Z_{pa}+Z_{aa})/4$, where $p/a$ means that the path integral is computed for Grassmann fields with periodic/anti-periodic boundary conditions along $x$ and $y$ directions.
Similarly, $z_{(e,o)} = (-Z_{pp}+Z_{pa}-Z_{ap}+Z_{aa})/4$, $z_{(o,e)}=(-Z_{pp}-Z_{pa}+Z_{ap}+Z_{aa})/4$, and $z_{(o,o)}=(-Z_{pp}+Z_{pa}+Z_{ap}-Z_{aa})/4$.

To compute dimer densities and dimer-dimer correlation functions in the
RK wave function, we
use the same path-integral representation of interacting Grassmann fields. The dimer density on a given link $(ij)$ is
\bea
\avg{D_{ij}} &\equiv& \frac{\bra{\Psi_0} D_{ij} \ket{\Psi_0}}{\braket{\Psi_0}{\Psi_0}},\nn\\
&=&\frac1{Z}\sgn(t_{ij})\frac{\delta Z}{\delta t_{ij}},
\eea
where $\sgn$ is the sign function which is used to make the density non-negative. The explicit expressions are slightly different between first neighbor and second neighbor links. For first neighbor links, the density density is given by
\bea
\avg{D_{ij}}=\frac{\sgn(t_{ij})}{Z}\int [da](-ia_i a_j)e^{-S}.
\eea
The density density on second neighbor links $(ij)$ is
\bea
&&\avg{D_{ij}}
=\frac{\sgn(t_{ij})}{Z}\nn\\
&&\times\int [da]\Big[(-ia_i a_j)(1-\sum_{kl\in\avg{ijkl}}it_{kl}a_k a_l)\Big]e^{-S}.
\eea

Similarly, we can obtain the dimer-dimer correlation functions from the path integral representation. The dimer-dimer correlation function is given by
\bea
\avg{D_{ij}D_{i'j'}}&\equiv& \frac{\bra{\Psi_0} D_{ij} D_{i'j'} \ket{\Psi_0}}{\braket{\Psi_0}{\Psi_0}},\nn\\
&=&\frac{1}{Z}\sgn(t_{ij}t_{i'j'})\frac{\delta^2 Z}{\delta t_{ij}\delta t_{i'j'}},
\eea
for which an explicit expression in terms of the Grassmann path integral can be written out  straightforwardly.

\begin{thebibliography}{33}
\bibitem{footnote}
Away from
    solvable points, the power-law prefactor may change to a $(2+z)$-dimensional Ornstein-Zernicke form.
\end{thebibliography}
\end{document}